\documentclass{iau379}

\usepackage{graphicx}
\usepackage{amsmath}
\usepackage{multirow}

\begin{document}

\lefttitle{R. A. N. Brooks et al.}
\righttitle{IAU Symposium 379: North-south asymmetry of ALFALFA H\,{\textsc{i}}\,WF}

\jnlPage{1}{7}
\jnlDoiYr{2023}
\doival{10.1017/xxxxx}

\aopheadtitle{Proceedings of IAU Symposium 379}
\editors{P. Bonifacio,  M.-R. Cioni, F. Hammer, M. Pawlowski, and S. Taibi, eds.}

\title{Origins of the north-south asymmetry in the ALFALFA H\,{\textsc{i}} velocity width function}

\author{
Richard A. N. Brooks,$^{1,2,3}$
Kyle A. Oman,$^{2,3}$
Carlos. S. Frenk $^{2,3}$}\affiliation{$^{1}$Department of Physics and Astronomy, University College London, London WC1E 6BT, UK\\
$^{2}$Institute for Computational Cosmology, Durham University, South Road, Durham DH1 3LE, UK\\
$^{3}$Department of Physics, Durham University, South Road, Durham DH1 3LE, UK\\}

\begin{abstract}
The number density of extragalactic {21-cm} radio sources as a function of their spectral line-widths -- the H\,\textsc{i} width function (H\,\textsc{i}\,WF) -- is a tracer of the dark matter halo mass function. The ALFALFA {21-cm} survey measured the H\,\textsc{i}\,WF in northern and southern Galactic fields finding a systematically higher number density in the north; an asymmetry which is in tension with $\Lambda$ cold dark matter models which predicts the H\,\textsc{i}\,WF should be identical everywhere if sampled in sufficiently large volumes. We use the \emph{Sibelius-DARK} N-body simulation and semi-analytical galaxy formation model \emph{GALFORM} to create mock ALFALFA surveys to investigate survey systematics. We find the asymmetry has two origins: the sensitivity of the survey is different in the two fields, and the algorithm used for completeness corrections does not fully account for biases arising from spatial galaxy clustering. Once survey systematics are corrected, cosmological models can be tested against the H\,\textsc{i}\,WF. 
\end{abstract}


\maketitle

\section{Introduction}\label{sec:introduction}

The kinematics of observable tracers in a galaxy are linked to the HMF because the maximum circular velocity of a halo, $v_\mathrm{max}$, is correlated to its mass \citep{1997ApJ...490..493N}. Connecting a kinematic tracer to $v_\mathrm{max}$ often requires additional modelling, e.g., {21-cm} spectral line widths. The {21-cm} velocity spectrum is the H\,\textsc{i} mass-weighted line-of-sight velocity distribution, the full width at half maximum of the spectrum thus parameterises the line-width, $w_{50}$. Instead of attempting to infer $v_\mathrm{max}$ from line-width measurements, a more straightforward technique is to predict the number density of extragalactic {21-cm} sources as a function of $w_{50}$ -- the H\,\textsc{i} width function (H\,\textsc{i}\,WF) -- and then compare with observational measurements. 

\cite{2009ApJ...700.1779Z}, and more recently \citet{2022MNRAS.509.3268O}, found differences in the normalisation of the ALFALFA H\,\textsc{i}\,WF when measured in its northern and southern Galactic field. The driver of this asymmetry has been previously only speculated \citep[e.g., see][]{2022MNRAS.509.3268O}. We use the \emph{Sibelius-DARK} N-body simulation \citep{2022MNRAS.512.5823M} populated with galaxies using the \emph{GALFORM} semi-analytical model \citep{2016MNRAS.462.3854L} to create mock ALFALFA surveys. Our mock surveys provide a footing for an investigation into survey systematics that may be responsible for driving the asymmetry; our approach enables us to provide the first quantitative estimates for the magnitude of these effects and assess whether they can explain the observed asymmetry. A full account of this work is detailed in \citet{2023MNRAS.tmp.1203B}.

\section{Mock {21-cm} survey: \emph{Sibelius-DARK} and \emph{GALFORM}}\label{sec:methods}

The ALFALFA survey \citep{2005AJ....130.2598G} mapped $\sim 7,000 \deg^2$ of the sky visible from Arecibo at {21-cm} wavelengths out to $\sim250\,\mathrm{Mpc}$. The survey was completed in 2012, and is composed of two separate fields on the sky; one in the northern Galactic hemisphere, visible during the spring, and the other in the southern Galactic hemisphere, visible during the autumn. By convention, these fields are labelled ‘spring' and ‘fall', respectively \citep{2018MNRAS.477....2J}. We define a selection of ALFALFA sources from which we measure the H\,\textsc{i}\,WF in a similar way to \citet{2022MNRAS.509.3268O}, but we impose a distance cut $\mathrm{d_{mw} \leq 200\,\mathrm{Mpc}}$ in order to facilitate comparison with the \emph{Sibelius-DARK} simulation which is contaminated by low-resolution particles from outside of the zoom-in region beyond this distance \citep{2022MNRAS.512.5823M}. Only sources above the $50$~per~cent completeness limit (CL) of the survey are selected. There are $20,857$ sources above the $50$~per~cent CL in the ALFALFA catalogue, of which $13,006$ are in the spring field and $7,851$ are in the fall field. 

‘Simulations Beyond The Local Universe' \citep[\emph{Sibelius};][]{2022MNRAS.509.1432S} aims to connect the Local Group with its cosmological environment using $\Lambda$CDM initial conditions that are constrained such that the large-scale structure is accurately reproduced. The first simulation from the \emph{Sibelius} project is \emph{Sibelius-DARK} \citep{2022MNRAS.512.5823M}, a realisation of a volume constrained within $200\,\mathrm{Mpc}$ of the Milky Way,  making it ideal to compare with the ALFALFA survey. The N-body \emph{Sibelius-DARK} simulation has its galaxy population inferred using the \citet{2016MNRAS.462.3854L} variant of the semi-analytical model of galaxy formation \emph{GALFORM}. 

Four steps are required to construct a mock {21-cm} survey from the simulation: calculation of the galactic circular velocity curve, determination of the amount of H\,{\textsc{i}} gas as a function of line-of-sight velocity to produce a {21-cm} spectrum, convolution with a kernel to model the thermal broadening of the {21-cm} line, and application of the selection criteria consistent with the ALFALFA survey. The result is a mock survey containing $21,031$ sources, of which $12,408$ are in the spring field and $8,623$ are in the fall field. 


Finally, to account for the undetected galaxy population we use the $1/V_\mathrm{eff}$ maximum-likelihood estimator. The $1/V_\mathrm{eff}$ estimator is 2D because the CL of the ALFALFA survey depends both on the integrated {21-cm} flux and line-width of the source. From the summation of the values of the $1/V_\mathrm{eff}$ weights in 2D bins in $M_{\mathrm{HI}}$ and $w_{50}$ we compute the 2D H\,\textsc{i}\, mass-width function. The sum along the $w_{50}$ axis gives the H\,\textsc{i}\,MF, while the same along the mass axis gives the H\,\textsc{i}\,WF.

\section{\texorpdfstring{ALFALFA and \emph{Sibelius-DARK} + \emph{GALFORM} H\,\textsc{i}\,WFs}{ALFALFA and Sibelius-DARK + GALFORM HI width functions}}\label{sec:results}

\begin{figure*}
    \centering
	\includegraphics[width=0.85\textwidth]{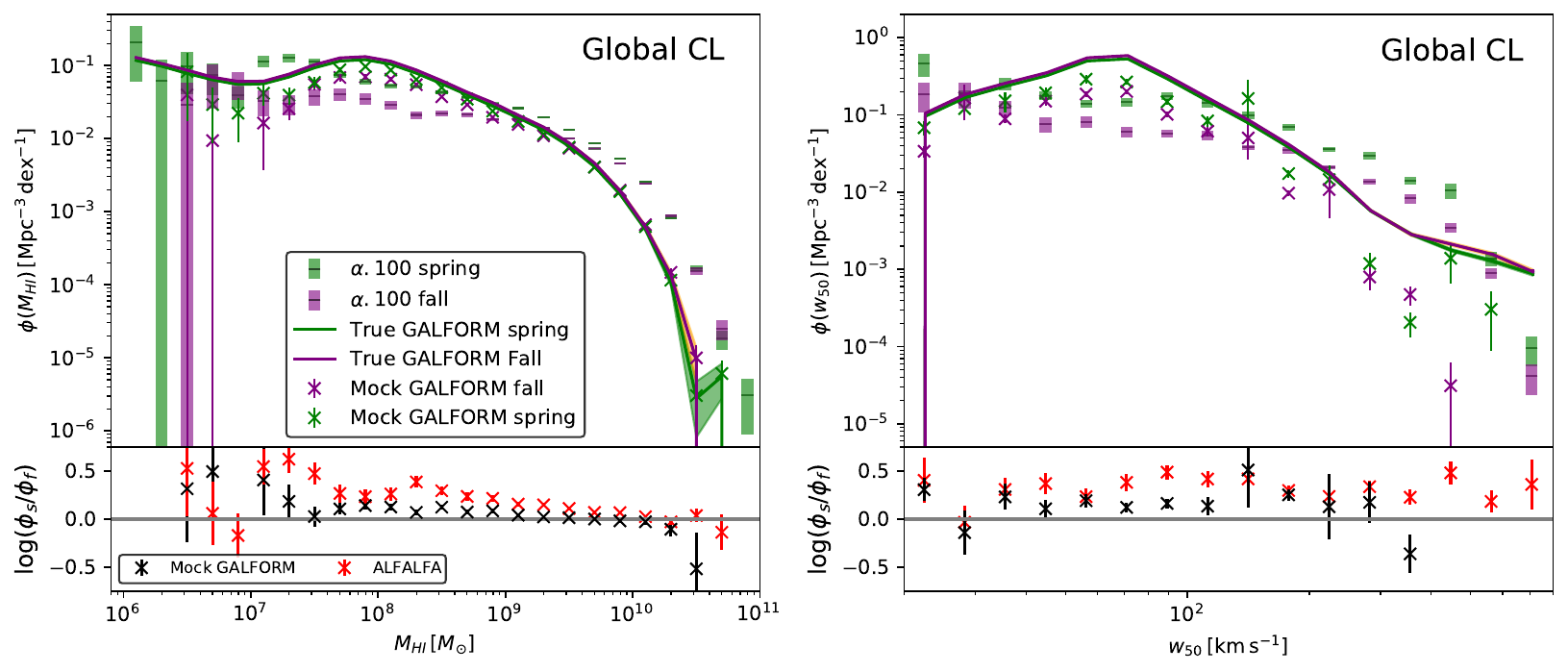}
    \includegraphics[width=0.85\textwidth]{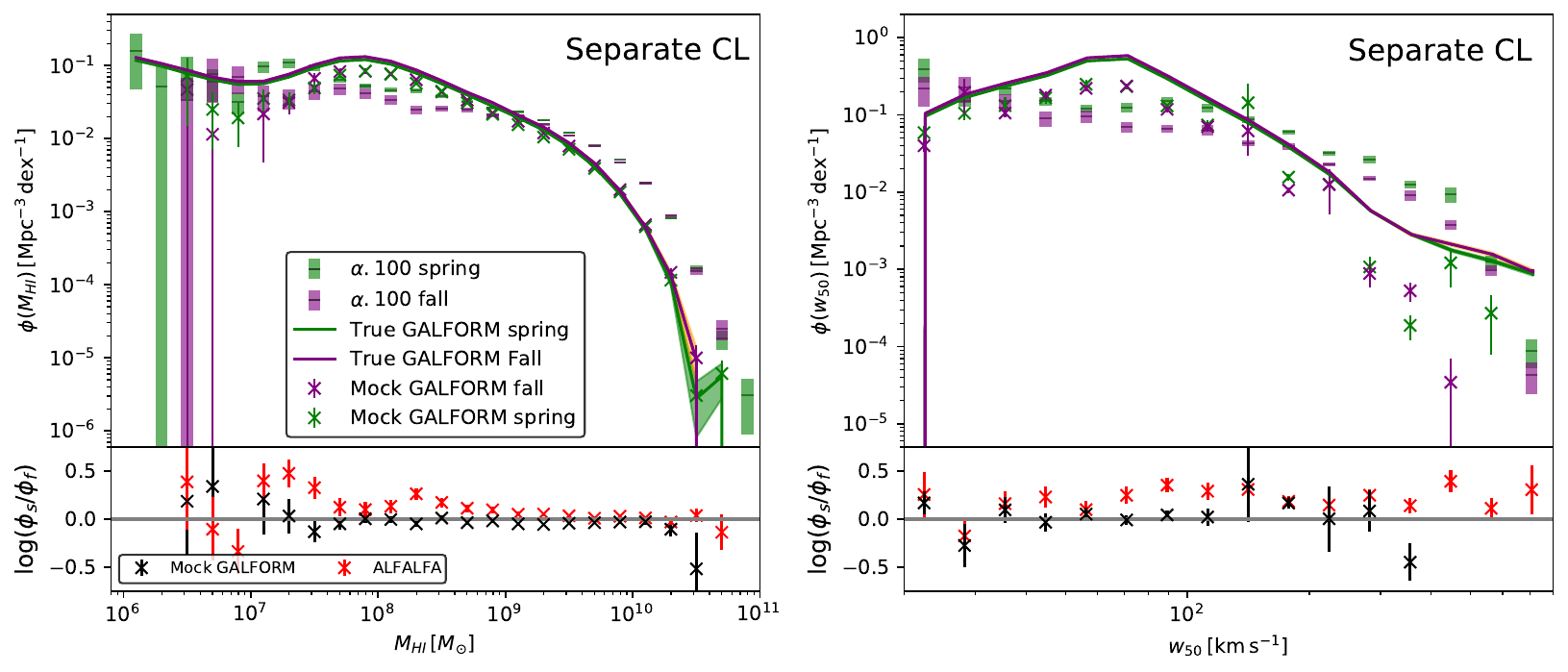}
    \caption{\emph{Upper left panel:} the H\,\textsc{i} mass function (H\,\textsc{i}\,MF) calculated by adopting the same completeness limit in the spring and fall fields. The main panel shows the H\,\textsc{i}\,MF measured from the ALFALFA ($\alpha.100$, dashes with shaded 1$\sigma$ uncertainties) and \emph{GALFORM} catalogues (crosses with 1$\sigma$ uncertainties shown with error bars), separately for the two ALFALFA fields, spring (green) and fall (purple). Additionally, the ‘true' \emph{GALFORM} H\,\textsc{i}\,MF (solid lines with 1$\sigma$ uncertainty shown with shaded band) is shown for the spring and fall fields. The lower sub-panel shows the ratio of the spring and fall \emph{GALFORM} (black crosses) and ALFALFA (red crosses) H\,\textsc{i}\,MFs with 1$\sigma$ uncertainties. \emph{Upper right panel:}  The H\,\textsc{i} width function (H\,\textsc{i}\,WF) calculated by adopting the same completeness limit in the spring and fall fields. Lines and symbols are as in upper left panel. \emph{Lower left and right panels:} similar to upper panels, but showing the H\,\textsc{i}\,MF and H\,\textsc{i}\,WF calculated by adopting separate completeness limits in the spring and fall fields. }
    \label{fig:HIMF_HIWF}
\end{figure*}

Understanding the origins of the asymmetric H\,\textsc{i}\,WFs is crucially important in the context of using it as a constraint on cosmology. The $\Lambda$CDM cosmological model predicts that the dark matter HMF should be universal \citep[in shape and normalisation, e.g.][]{1988ApJ...327..507F, 1996MNRAS.282..347M, 2002MNRAS.329...61S, 2009MNRAS.399.1773C} and therefore similar in the two fields surveyed in ALFALFA because the volumes sampled are sufficiently large. We have checked this explicitly in the \emph{Sibelius-DARK} simulation: the HMFs in the spring and fall volumes differ by no more than $8$~per~cent (within their uncertainties) at any halo mass $10^{8}<M_{\mathrm{vir}}/\mathrm{M}_\odot<10^{14}$. The most straightforward prediction for the H\,\textsc{i}\,WF is that it should also have the same shape and normalisation (within about 8~per~cent) in the two fields. In the right panels of Fig.~\ref{fig:HIMF_HIWF} we show the simulation H\,\textsc{i}\,WF of all galaxies with $M_{\mathrm{HI}}>10^{6}\,\mathrm{M}_{\odot}$ in the spring and fall survey fields (regardless of whether they would be detected) with the green and purple solid lines, respectively. The two H\,\textsc{i}\,WF curves differ by no more than $8$~per~cent. If asymmetry of the H\,\textsc{i}\,WFs cannot otherwise be explained, then the $\Lambda$CDM model could be called into question. 

First, we make the measurement of the H\,\textsc{i}\,MF and H\,\textsc{i}\,WF for the spring and fall mock catalogues separately, following the procedure outlined in \citet{2022MNRAS.509.3268O} where the globally derived CL for the ALFALFA survey is assumed in the $1/V_\mathrm{{eff}}$ estimator; top-left and right panels of Fig.~\ref{fig:HIMF_HIWF} respectively show the H\,\textsc{i}\,MFs and H\,\textsc{i}\,WFs measured using this approach. The measurements of \citet[][]{2022MNRAS.509.3268O}, now retaining only sources within $200\,\mathrm{Mpc}$, are also shown in these panels for comparison. We repeat the measurement of the H\,\textsc{i}\,MF and H\,\textsc{i}\,WF for the spring and fall ALFALFA and mock catalogues, but this time assuming CLs derived separately for the spring and fall ALFALFA fields in the calculation of the $1/V_{\mathrm{eff}}$ weights; bottom panels of Fig.~\ref{fig:HIMF_HIWF}. By adopting the separately derived CLS, the spring and fall mock survey H\,\textsc{i}\,WFs are no longer systematically offset from each other. Therefore, when the ‘correct' CLs used to construct the mock catalogues are also used to correct them for incompleteness, the $1/V_\mathrm{eff}$ algorithm correctly recovers the fact that the H\,\textsc{i}\,WFs in the two fields are indistinguishable. For ALFALFA, the systematic offset between the spring and fall H\,\textsc{i}\,WFs is reduced when the ‘correct' CLs are used in the $1/V_\mathrm{eff}$ estimator, but does not disappear: the median ratio between the spring to fall H\,\textsc{i}\,WFs is $\log_{10}(\phi_{s}\,/\,\phi_{f}) = 0.25 \pm 0.09$. The persisting asymmetry suggests that this is not the only systematic effect influencing the measurement.

\begin{figure}
    \centering
    \includegraphics[width=0.46\textwidth]{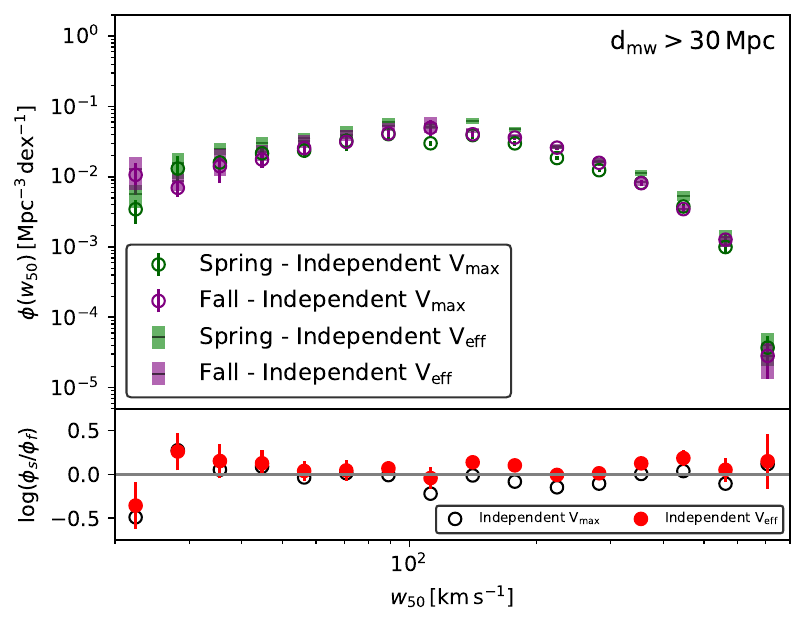}
    \caption{The spring (green) and fall (purple) H\,\textsc{i}\,WF  derived using the $1/\mathrm{V_{eff}}$ (dashes) and $1/\mathrm{V_{max}}$ (open circles) weights assigned independently but including only those sources detected at distances $d_{\mathrm{mw}} > 30\,\mathrm{Mpc}$. The lower sub panel again shows the spring-to-fall ratio for the independent $1/\mathrm{V_{eff}}$ (red points) and independent $1/\mathrm{V_{max}}$ (open black circles) $1/\mathrm{V_{eff}}$ weights.} 
    \label{fig:AA_HIWF_losclustering}
\end{figure}

The origin of the remaining asymmetry stems from the number of detected sources in the foreground of the survey ($d_{\mathrm{mw}}\lesssim 30\,\mathrm{Mpc}$). The $1/\mathrm{V_{eff}}$ estimator incorrectly extrapolates the foreground overdensity in the spring field through the entire survey volume, biasing the spring field to higher number densities. This effect is similar to what would occur if we used the $1/V_\mathrm{max}$ estimator \citep{1968ApJ...151..393S}, which assumes that galaxies are uniformly distributed in space, but less severe -- the $1/V_{\mathrm{eff}}$ algorithm is intended to compensate for non-uniformity in the galaxy distribution, but does so imperfectly. In Fig.~\ref{fig:AA_HIWF_losclustering}, we show the H\,\textsc{i}\,WFs in the two fields measured with both (i)~a conservative CL imposed to remove bias from the differing sensitivity in the two survey field and (ii) sources within $30\,\mathrm{Mpc}$ removed. In this case the H\,\textsc{i}\,WFs in the two fields are very close to agreement; the median ratio between the spring to fall H\,\textsc{i}\,WFs derived using the independent $V_\mathrm{eff}$ weights is $\log_{10}(\phi_{s}\,/\,\phi_{f}) = 0.07 \pm 0.12$. 

In summary, we attribute the asymmetry between the spring and fall ALFALFA H\,\textsc{i}\,WFs to: (i)~the adopted completeness limit for the survey; and (ii)~the $1/V_\mathrm{eff}$ estimator incorrectly extrapolating the foreground overdensity in the spring field through the entire survey volume. Accounting for these systematic effects leads to H\,\textsc{i}\,WFs in the spring and fall fields that are consistent with being identical, as expected from $\Lambda$CDM cosmologies.

\section{\texorpdfstring{Outlook -- the H\,\textsc{i}\,WF as a cosmological constraint}{Summary and outlook -- the HIWF as a cosmological constraint}}\label{sec:outlook}


The influence of individual over/underdense regions on the calculation of $1/V_\mathrm{eff}$ weights can be mitigated by simply surveying a larger area on the sky; the ongoing WALLABY survey \citep{2020Ap&SS.365..118K} will cover an area about four times wider than that covered by ALFALFA. As the abundance of sources at low line-widths is sensitive to the adopted dark matter model, the H\,\textsc{i}\,WF has the potential to become a stringent test of cosmological models. Realising this potential will require a deeper understanding of the systematic biases influencing measurements, such as those due to spatial (or temporal) variability in survey sensitivity. Progress on theoretical issues is also needed. We look forward to the prospect of using the H\,\textsc{i}\,WF as a constraint on the nature of dark matter.

\setlength{\bibhang}{0pt}

\end{document}